\documentclass[english,aps,twocolumn, notitlepage, prl,longbibliography,superscriptaddress]{revtex4-1}
\usepackage[latin1]{inputenc}
\usepackage{color}
\usepackage{babel}
\usepackage{amsmath}
\usepackage{amssymb}
\usepackage{graphicx}
\usepackage[unicode=true,
 bookmarks=false,
 breaklinks=false,pdfborder={0 0 1},backref=false,colorlinks=true]
 {hyperref}
\hypersetup{
 citecolor=blue,linkcolor=magenta}

\makeatletter
\usepackage{braket}
\usepackage{dsfont}
\usepackage[caption=false]{subfig}

\newcommand{\AR}[1]{{\color{black} #1}}
\newcommand{\CC}[1]{{\color{black} #1}}

\makeatother

\begin{document}

\title{Learning hard quantum distributions with variational autoencoders}

\author{Andrea Rocchetto}
\email{andrea.rocchetto@spc.ox.ac.uk}
\affiliation{Department of Computer Science, University of Oxford, Oxford OX1 3QD, UK}
\affiliation{Department of Materials, University of Oxford, Oxford OX1 3PH, UK}
\affiliation{Department of Computer Science, University College London, London 
WC1E 6EA, UK}

\author{Edward Grant}
\email{edward.grant.16@ucl.ac.uk}

\affiliation{Department of Computer Science, University College London, London 
WC1E 6EA, UK}

\author{Sergii Strelchuk}

\affiliation{DAMTP, University of Cambridge, Cambridge CB3 0WA, UK}

\author{Giuseppe Carleo}
\affiliation{Institute for Theoretical Physics, ETH Z\"urich, Z\"urich 8093, Switzerland}
\affiliation{Center for Computational Quantum Physics, Flatiron Institute, New York 10010, USA}

\author{Simone Severini}

\affiliation{Department of Computer Science, University College London, London 
WC1E 6EA, UK}

\affiliation{Institute of Natural Sciences, Shanghai Jiao Tong University, Shanghai, China}
\begin{abstract}
The exact description of many-body quantum systems represents one
of the major challenges in modern physics, because it requires an
amount of computational resources that scales exponentially with the
size of the system. Simulating the evolution of a state, or even storing
its description, rapidly becomes intractable for exact classical algorithms.
Recently, machine learning techniques, in the form of restricted Boltzmann
machines, have been proposed as a way to efficiently represent certain
quantum states with applications in state tomography and ground state
estimation. Here, we introduce a practically usable deep architecture
for representing and sampling from probability distributions of quantum
states. Our representation is based on variational auto-encoders,
a type of generative model in the form of a neural network. We show
that this model is able to learn efficient representations of states
that are easy to simulate classically and can compress states that
are not classically tractable. Specifically, we consider the learnability
of a class of quantum states introduced by Fefferman and Umans. Such
states are provably hard to sample for classical computers, but not
for quantum ones, under plausible computational complexity assumptions.
The good level of compression achieved for hard states suggests these
methods can be suitable for characterizing states of the size expected
in first generation quantum hardware. 
\end{abstract}
\maketitle

\section*{Introduction}

One of the most fundamental tenets of quantum physics is that the
physical state of a many-body quantum system is fully specified by
a high-dimensional function of the quantum numbers, the wave-function.
As the size of the system grows the number of parameters required
for its description scales exponentially in the number of its constituents.
This complexity is a severe fundamental bottleneck in the numerical
simulation of interacting quantum systems. Nonetheless, several approximate
methods can handle the exponential complexity of the wave function
in special cases. For example, \textit{quantum Monte Carlo} methods (QMC),
allow to sample exactly from many-body states free of sign problem~\cite{nightingale1998quantum,gubernatis2016quantum,suzuki1993quantum},
and \textit{Tensor Network approaches} (TN), very efficiently represent low-dimensional
states satisfying the area law for entanglement \cite{verstraete2008matrix,orus2014practical}.

Recently, machine learning methods have been introduced to tackle
a variety of tasks in quantum information processing that involve
the manipulation of quantum states. These techniques offer greater
flexibility and, potentially, better performance, with respect to
the methods traditionally used. Research efforts have focused on representing
quantum states in terms of \textit{restricted Boltzmann machines}
(RBMs). The RBM representation of the wave function, introduced by
Carleo and Troyer~\cite{carleo2017solving}, has been successfully
applied to a variety of physical problems, ranging from strongly correlated
spins \cite{carleo2017solving,deng_quantum_2017}, and fermions \cite{nomura_restricted-boltzmann-machine_2017}
to topological phases of matter \cite{deng_exact_2016,glasser_neural_2017,kaubruegger_chiral_2017}.
Particularly relevant to our purposes is the work by Torlai \textit{et
al.}~\cite{torlai2017many} that makes use of RBMs to perform \textit{quantum
state tomography} of states whose evolution can be simulated in polynomial
time using classical methods (\textit{e.g.} \textit{matrix product
states} (MPS)~\cite{perez2006matrix}). Although it is remarkable
that RBMs can learn an efficient representation of this class of states
without any explicitly programmed instruction, it remains unclear
how the model behaves on states where no efficient classical description
is available.

Theoretical analysis of the representational power of RBMs has been
conducted in a series of works \cite{gao2017efficient,chen_equivalence_2017,huang_neural_2017,deng_quantum_2017,clark_unifying_2017}.
Gao and Duan, in particular, showed that RBMs cannot efficiently encode
every quantum state \cite{gao2017efficient}. They proved that Deep
Boltzmann Machines (DBMs) with complex weights, a multilayer variant
of RBMs, can efficiently represent most physical states. Although
this result is of great theoretical interest the practical application
of complex-valued DBMs in the context of unsupervised learning has
not yet been demonstrated due to a lack of efficient methods to sample
efficiently from DBMs when the weights are complex-valued. The absence
of practically usable deep architectures remains an important limitation
of current neural network based learning methods for quantum systems.
Indeed, several research efforts on neural networks~\cite{mhaskar2016learning,telgarsky2016benefits,eldan2016power}
have shown that depth significantly improves the representational
capability of networks for some classes of functions (such as compositional
functions).

In this Paper, we address several open questions with neural network
quantum states. First, we study how the depth of the network affects
the ability to compress quantum many-body states. This task is achieved
upon introduction of a deep neural network architecture for encoding
probability distribution of quantum states, based on \textit{variational
autoencoders} (VAEs)~\cite{kingma2013auto}. We benchmark the performance
of deep networks on states where no efficient classical description
is known, finding that depth systematically improves the quality of
the reconstruction for states that are computationally tractable and
for hard states that can be efficiently constructed with a quantum
computer. Surprisingly, the same does not apply for hard states that
cannot be efficiently constructed by means of a quantum process. Here,
depth does not improve the reconstruction accuracy. 

Second, we show that VAEs can learn efficient representations of computationally
tractable states and can reduce the number of parameters required
to represent an hard quantum state up to a factor $5$. This improvement
makes VAE states a promising tool for the characterization of early
quantum devices that are expected to have a number of qubits that
is slightly larger than what can be efficiently simulated using existing
methods~\cite{boixo2016characterizing}. 

\subsection*{Encoding quantum probability distributions with VAEs}

\label{sec:qVAE}

Variational autoencoders (VAEs), introduced by Kingma and Welling
in 2013~\cite{kingma2013auto}, are generative models based on layered
neural networks. Given a set of i.i.d. data points $X=\{x^{(i)}\}$,
where $x^{(i)}\in\mathbb{R}^{n}$, generated from some distribution
$p_{\theta}(x^{(i)}|z)$ over Gaussian distributed latent variables
$z$ and model parameters $\theta$, finding the posterior density
$p_{\theta}(z|x^{(i)})$ is often intractable. VAEs allow for approximating
the true posterior distribution, with a tractable approximate model
$q_{\phi}(z|x^{(i)})$, with parameters $\phi$, and provide an efficient
procedure to sample efficiently from $p_{\theta}(x^{(i)}|z)$. The
procedure does not employ Monte Carlo methods.

\begin{figure}
\begin{centering}
\includegraphics[scale=0.35]{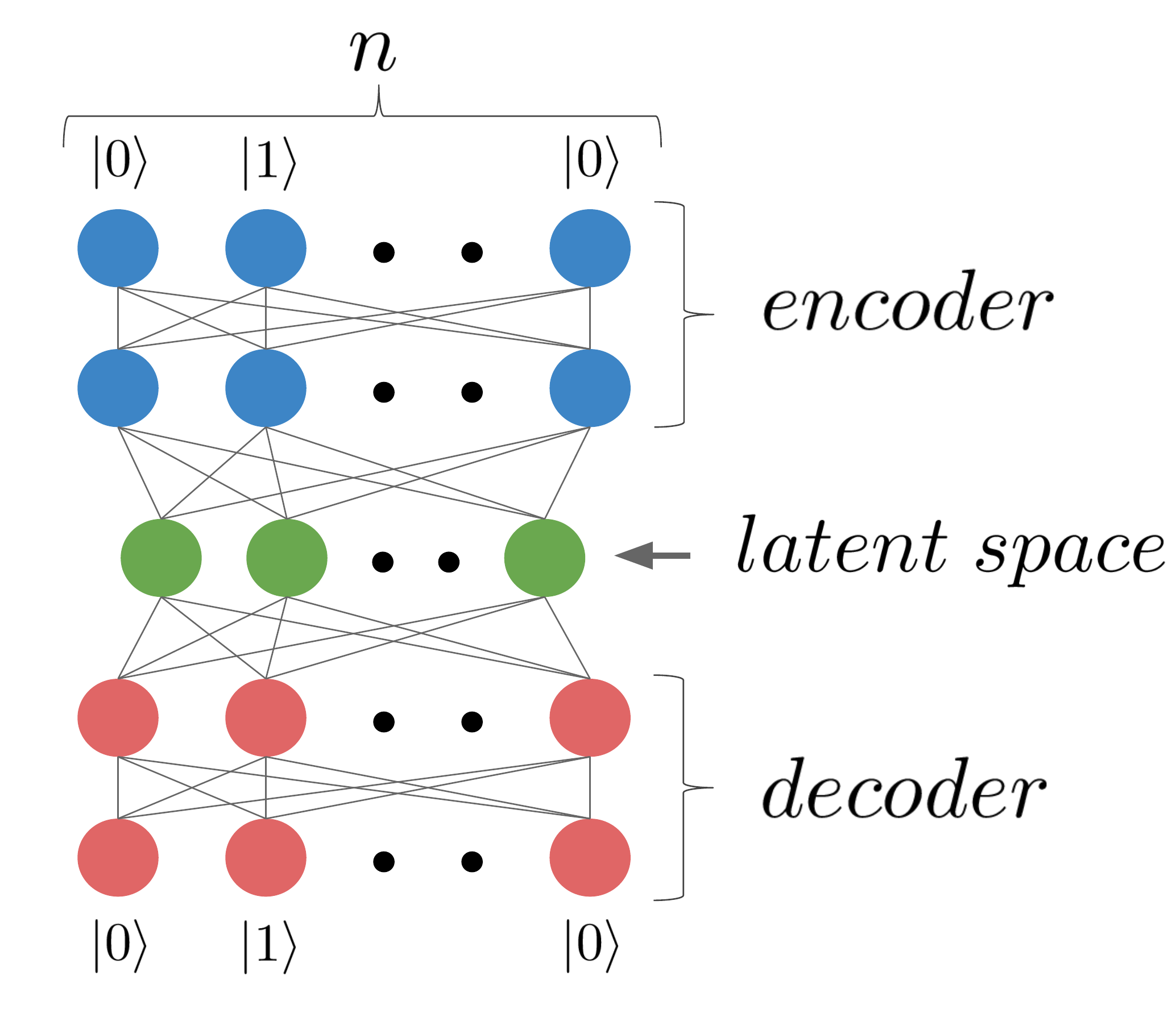}
\par\end{centering}
\caption{\label{fig:VAE_schematic}\textbf{Encoding quantum probability distributions
with VAEs.} A VAE can be used to encode and then generate samples
according to the probability distribution of a quantum state. Each
dot corresponds to a neuron and neurons are arranged in layers. Input
(top), latent, and output (bottom) layers contain $n$ neurons. The
number of neurons in the other layers is a function of the compression
and the depth. Layers are fully connected with each other with no
intra layer connectivity. The network has three main components: the
encoder (blue neurons), the latent space (green), and the decoder
(red). Each edge of the network is labelled by a weight $\theta$.
The total number of weights $m$ in the decoder corresponds to the
number of parameters used to represent a quantum state. The network
can approximate quantum states using $m<2^{n}$ parameters. The model
is trained using a dataset consisting of basis elements drawn according
to the probability distribution of a quantum state. Elements of the
basis are presented to the input layer on top of the encoder and,
during the training phase, the weights of the network are optimized
in order to reconstruct the same basis element in the output layer.}
\end{figure}

As shown in Fig.~\ref{fig:VAE_schematic} a VAE is composed of three
main components. The \textit{encoder} that is used to project the
input in the \textit{latent space} and the \textit{decoder} that is
used to reconstruct the input from the latent representation. Once
the network is trained the encoder can be dropped and, by generating
samples in the latent space, it is possible to sample according to
the original distribution. In graph theoretic terms, the graph representing
a network with a given number of layers is a \textit{blow up} of a
directed path on the same number of vertices. Such a graph is obtained
by replacing each vertex of the path with an independent set of arbitrary
but fixed size. The independent sets are then connected to form complete
bipartite graphs.

The model is trained by minimizing over $\theta$ and $\phi$ the
cost function: 
\begin{multline}
J(\theta,\phi,x^{(i)})=-\mathds{E}_{z\sim{q_{\phi}(z|x^{(i)})}}[\log p_{\theta}(x^{(i)}|z)]\\
+D_{\mathrm{KL}}(q_{\phi}(z|x^{(i)})||p_{\theta}(z))).\label{eq:jtheta}
\end{multline}
The first term (reconstruction loss) $-\mathds{E}_{z\sim{q_{\phi}(z|x^{(i)})}}[\log p_{\theta}(x^{(i)}|z)]$
is the expected negative log-likelihood of the $i$-th data-point
and favors choices of $\theta$ and $\phi$ that lead to more faithful
reconstructions of the input. The second term (regularization loss)
$D_{\mathrm{KL}}(q_{\phi}(z|x^{(i)})||p_{\theta}(z)))$ is the Kullback-Leibler
divergence between the encoder's distribution $q_{\phi}(z|x^{(i)})$
and the Gaussian prior on $z$. A full treatment and derivations of
the variational objective are given in~\cite{kingma2013auto}.

VAEs can be used to encode the probability distribution associated
to a quantum state. Let us consider an $n$-qubit quantum state $\ket{\psi}$,
with respect to a basis $\{\ket{b_{i}}\}_{i=1,\dots,2^{n}}$. We can
write the probability distribution corresponding to $\ket{\psi}$
as $p(b_{i})=|\langle b_{i}|\psi\rangle|^{2}$. If we consider the
computational basis, we can write $\ket{\psi}=\sum_{i=1}^{2^{n}}\psi_{i}\ket{i}$,
where each basis element corresponds to an $n$-bit string. A VAE
can be trained to generate basis elements $\ket{i}$ according to
the probability $p(i)=|\langle i|\psi\rangle|^{2}=|\psi_{i}|^{2}$.

We note that, in principle, it is possible to encode a full quantum
state (phase included) in a VAE. This requires samples taken from
more than one basis and a network structure that can distinguish among
the different inputs. The development of VAE encodings for full quantum
states will be left to future work.

We approximate the true posterior distribution across measurement
outcomes in the latent space $z$ with a multivariate Gaussian, having
diagonal covariance structure, zero mean and unit standard deviation.
The training set consists of a set of basis elements generated according
to the distribution associated with a quantum state. Following training,
the variables $z$ are sampled from a multivariate Gaussian and
used as the input to the decoder. By taking samples from this Gaussian
as input, the decoder is able to generate strings corresponding to
measurement outcomes that closely follow the distribution of measurement
outcomes used to train the network.

\subsection*{Hard and easy quantum states}
\label{sec:hard}

\AR{In this section we introduce a method to classify quantum states based on the hardness of sampling their probability distribution in a given basis. This will be used to assess the power of deep neural network models at representing many-body wave-functions. 

We now proceed to define two concepts that will be frequently used throughout the paper and form the basis of our classification method: \textit{reconstruction accuracy} and \textit{compression}. Let $\rho$ and $\sigma$ be $n$--qubit quantum states.  We say that $\sigma$ is a good representation of 
$\rho$ if the fidelity $F = \mathrm{Tr}(\sqrt{\rho^{1/2}\sigma \rho^{1/2}}) \geq 1 - \epsilon$ for an $\epsilon>0$. This accuracy metric cannot be immediately applied to the analysis of VAEs, that can only encode the probability distribution associated to a state. We now show that the fidelity can expressed in terms of the probability distributions over a measurement that maximally distinguishes the two states. Let $E=\{E_i\}$ be a POVM measurement. Then, using a result by Fuchs and Caves~\cite{fuchs1994ensemble} we can write
\begin{equation}
\label{eq:fidelity}
F = \min_{E} \sum_i \sqrt{\mathrm{Tr}(E_i \rho)\mathrm{Tr}(E_i \sigma)},
\end{equation}
where the minimum is taken over all possible POVMs. Note that $p(i) = \mathrm{Tr}(E_i \rho)$ and $q(i) = \mathrm{Tr}(E_i \sigma)$ are the probabilities of measuring the state $\rho$ and $\sigma$, respectively, in outcome labelled by $i$ and $\sum_i \sqrt{p(i)q(i)}$ is the Bhattacharyya coefficient between the two distributions.

Using Eq.~\ref{eq:fidelity} we can relate the complexity of a state with the problem of estimating the fidelity $F$. This corresponds to the hardness of sampling the probability distribution $p(i) = \mathrm{Tr}(E'_i \rho)$, where $E'$ minimises Eq.~\ref{eq:fidelity} (here we assume that sampling from the approximating distribution $q(i)$ is at most as hard as sampling from $p(i)$).

Throughout the paper, unless where explicitly mentioned, we will work with states that have only positive, real entries in the computational basis. In this case, it is easy to see that the Bhattacharyya coefficient between the distributions reduces to the fidelity and, hence, measurements in the $Z$ basis minimises Eq~\ref{eq:fidelity}. 

We remark that, if it is not possible to find a POVM for which Eq.~\ref{eq:fidelity} is minimised it is always possible to use the standard formulation of the fidelity as a metric in the context of VAEs. This can be accomplished by making use of $3$ VAEs to encode the state $\sigma$ over $3$ different basis. By using standard tomographic techniques, like maximum likelihood, measurements in a complete basis can be used to reconstruct the full density matrix. 

In order to connect the above definition of state complexity with VAEs we introduce the \textit{compression factor}. Given an $n$-qubit state that is represented by a VAE with $m$ parameters in the decoder, the compression factor is $C = \frac{m}{2^{n}}$. We say that a state $\rho$ is \textit{exponentially compressible} if there exists a network that approximates $\rho$ with high accuracy using $m = O(\mathrm{poly}(n))$ parameters. 

Once a network is trained, the cost of generating a sample is proportional to the number of parameters in the network. In this sense the complexity of a state is parametrised by the number of parameters used by a neural network representation. Based on these observation we define \textit{easy states} those that can be represented with high accuracy and exponential compression and \textit{hard states} those that can be represented with high accuracy using at least $O(\mathrm{exp}(n))$ parameters.
The last category includes: 1) states
that can be efficiently sampled with a quantum computer, but are
conjectured to have no classical algorithm to do so; 2) states that
cannot be efficiently obtained on a quantum computer starting from
some fixed product input state (\textit{e.g.} random states).

Under this definition, states that
admit an efficient classical description (such as stabilizer states or MPS with low bond dimension) are easy, because we known that $O(\mathrm{poly}(n))$ parameters are sufficient to specify the state. Specifically, for the class of easy states we consider
separable states obtained by taking the tensor product of $n$ different
$1$-qubit random states. More formally, we consider states of the
form $\ket{\tau}=\bigotimes_{i=1}^{n}\ket{r_{i}}$ where $\ket{r_{i}}$
are random $1$-qubit states. These states can be described using
only $2n$ parameters. }

Among the class of hard states of the first kind, we study the learnability
of a type of hard distributions introduced in~\cite{fefferman2014power}
which can be sampled exactly on a quantum computer. These distributions
are conjectured to be hard to approximately sample from classically
\textendash{} the existence of an efficient sampler would lead to
the collapse of the Polynomial Hierarchy under some natural conjectures
described in~\cite{fefferman2014power,aaronson2011computational}.
We discuss how to generate this type of states in the Methods section.

Finally, for the second class of hard states, we consider random pure
states. These are generated by normalizing a $2^{n}$ dimensional
complex vector drawn from the unit sphere according to the Haar measure.

\begin{figure*}[t]   
\centering
\subfloat[]{\includegraphics[width=0.33\textwidth, keepaspectratio]{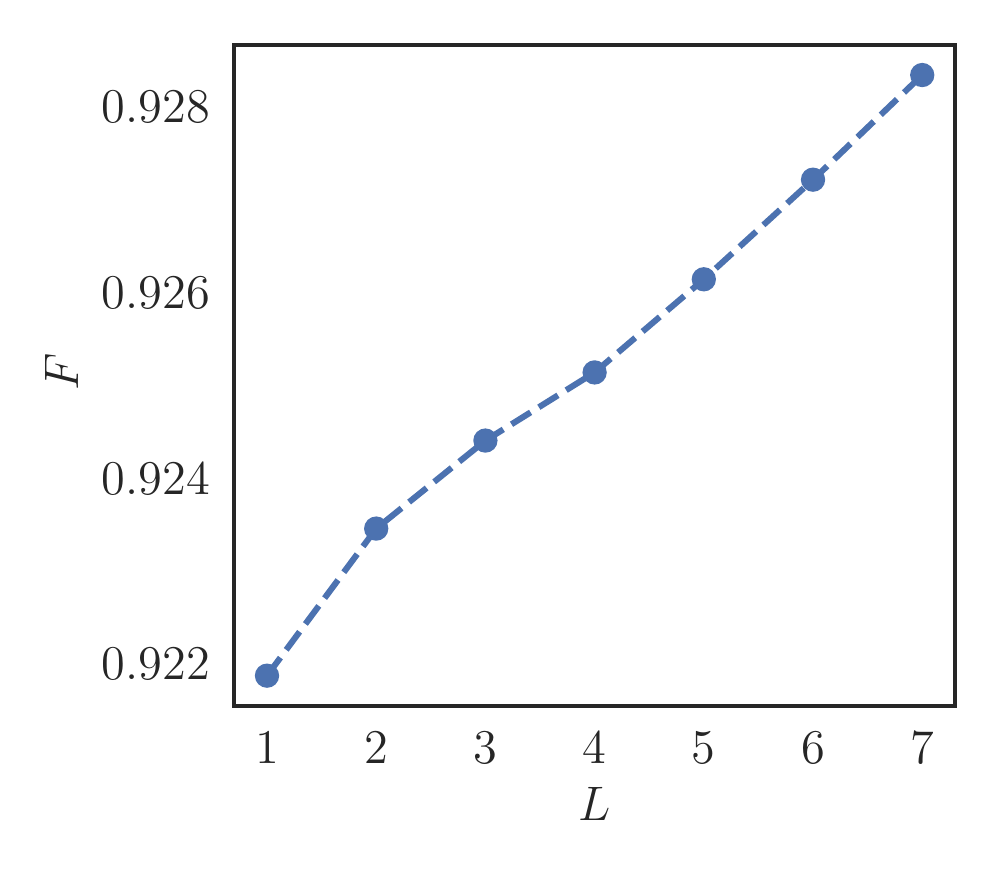}\label{fig:overlap_v_layers_hard}}
\subfloat[]{\includegraphics[width=0.33\textwidth, keepaspectratio]{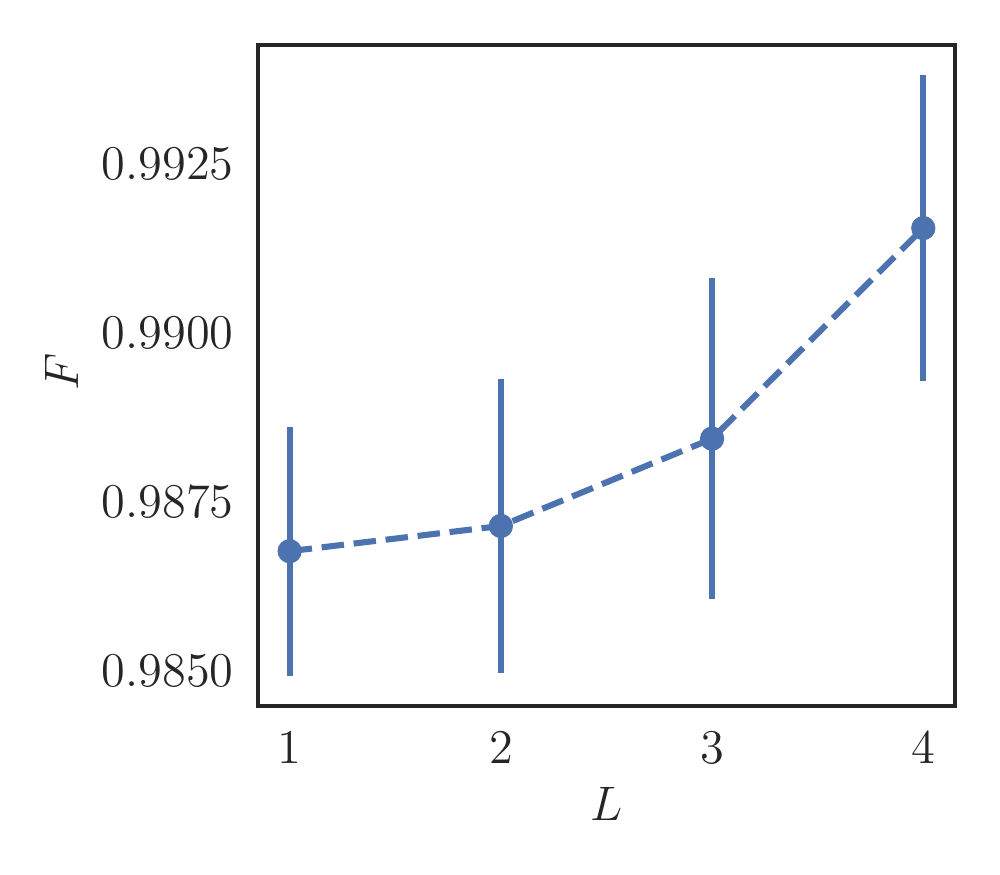}\label{fig:overlap_v_layers_product}}
\subfloat[]{\includegraphics[width=0.33\textwidth, keepaspectratio]{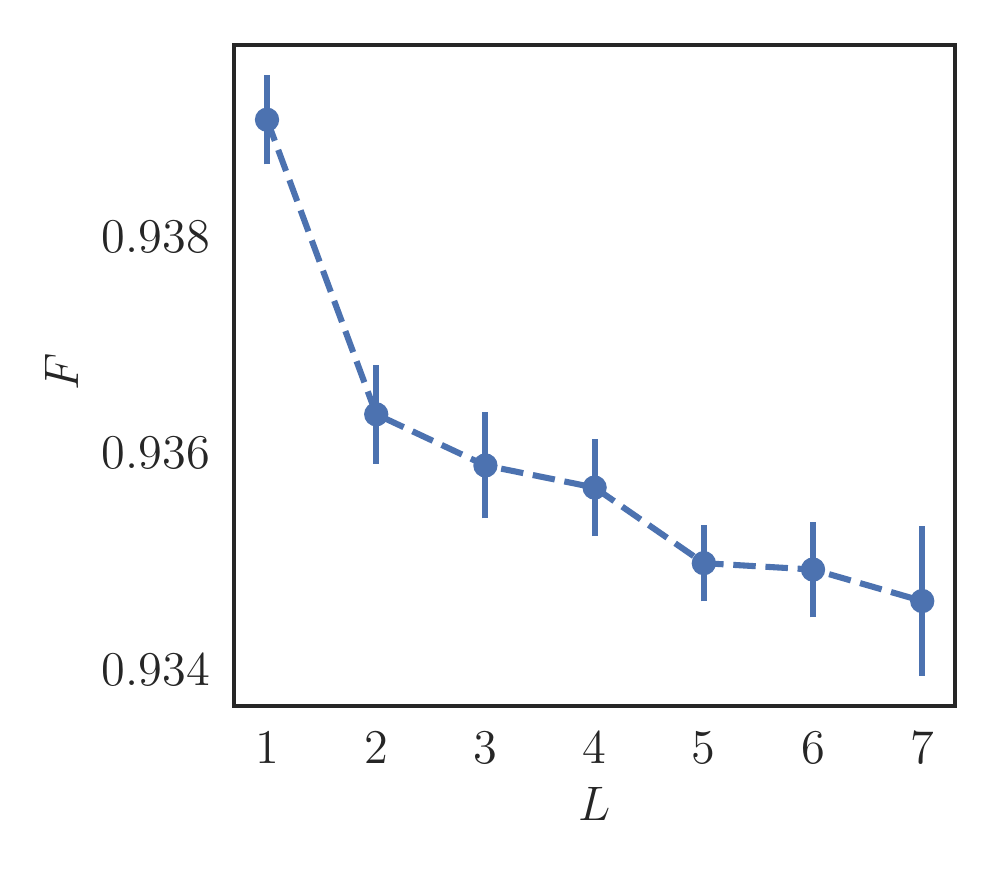}\label{fig:overlap_v_layers_random}}
\caption[Optional caption for list of figures 5-8]{\textbf{Depth affects the learnability of hard quantum states.} Fidelity
as a function of the number of layers in the VAE decoder for (a) an
$18$-qubit hard state that is easy to generate with a quantum computer,
(b) random $18$-qubit product states that admit efficient classical
descriptions and (c) random $15$-qubit pure states. Errors bars for
(b) and (c) show the standard deviation for an average of $5$ different
random states. The compression level $C$ is set to $C=0.5$ for (a)
and (c) and $C=0.015$ for (b) where $C$ is defined by $\frac{m}{2^{n}}$
where $m$ is the number of parameters in the VAE decoder and $n$
is the number of qubits. We use a lower compression rate for product
states because, due to their simple structure, even a $1$ layer network
achieves almost perfect overalp. Plot (b) makes use of up to 4 layers
in order to avoid the saturation effects discussed in the Methods
section.}
\label{fig:overlap_v_layers}
\end{figure*}

\section*{Results}

\subsection*{The role of depth in compressibility}

\label{sec:depth}

Classically, depth is known to play a significant role in the representational
capability of a neural network. Recent results, such as the ones by
Mhaskar, Liao, and Poggio~\cite{mhaskar2016learning}, Telgarsky~\cite{telgarsky2016benefits},
and Eldan and Shamir~\cite{eldan2016power} showed that some classes
of functions can be approximated by deep networks with the same accuracy
as shallow networks but with exponentially less parameters.

The representational capability of networks that represent quantum
states remains largely unexplored. Some of the known results are only
based on empirical evidence and sometimes yield to unexpected results.
For example, Morningstar and Melko~\cite{morningstar2017deep} showed
that shallow networks are more efficient than deep ones when learning
the energy distribution of a $2$-dimensional Ising model.

In the context of the learnability of quantum states Gao and Duan~\cite{gao2017efficient}
proved that DBMs can efficiently represent some states that cannot
be efficiently represent by shallow networks (\textit{i.e.} states
generated by polynomial depth circuits or $k$-local Hamiltonians
with polynomial size gap) \AR{using a polynomial number of hidden units}. However, there are no known methods to
sample efficiently from DBMs when the weights include complex-valued
coefficients.

We benchmark with numerical simulations the role played by depth in
compressing states of different levels of complexities. We focus on
three different states: an easy state (the completely separable state
discussed in the previous section), a hard state (according to
Fefferman and Umans), and a random pure state.

Our results are presented in Fig.~\ref{fig:overlap_v_layers}. Here,
by keeping the number of parameters in the decoder constant, we determine
the reconstruction accuracy of networks with increasing depth. Remarkably,
depth affects the reconstruction accuracy of hard quantum states.
This might indicate that VAEs are able to capture correlations in
hard quantum states. As a sanity check we notice that the network
can learn correlations in random product states and that depth does
not affect the learnability of random states.

Our simulations suggest a further link between neural network and
quantum states. This topic has recently received the attention of
the community. Specifically, Levine~\textit{et al.}~\cite{levine2017deep}
demonstrated that convolutional rectifier networks with product pooling
can be described as tensor networks. By making use graph theoretic
tools they showed that nodes in different layers model correlations
across different scales and that adding more nodes to deeper layers
of a network can make it better at representing non-local correlations.

\subsection*{Efficient compression of physical states}

\label{sec:probing}

In this section we focus our attention onto two questions: can VAEs
find efficient representations of easy states? What level of compression
can we obtain for hard states? Through numerical simulations we show
that VAEs can learn to efficiently represent some easy states (that
are challenging for standard methods) and achieve good levels of compressions
for hard states. Remarkably, our methods allow to compress up to a
factor $5$ the hard quantum states introduced in~\cite{fefferman2015power}.
We remark that the exponential hardness cannot be overcome for general
quantum states and our methods achieve only a factor improvement on
the overall complexity. This may nevertheless be sufficient to be
used as a characterisation tool where full classical simulation is
not feasible.

We test the performance of the VAE representation on two classes of
states: the hard states that can be constructed efficiently with a
quantum computer introduced by Fefferman and Umans~\cite{fefferman2015power}
and states that can be generated with a long-range Hamiltonian dynamics,
as found for example in experiments with ultra-cold ions~\cite{richerme2014non}.
The states generated through this evolution are highly symmetric physical
states. However, due to the bond dimension increasing exponentially
with the evolution time, these states are particularly challenging
for MPS methods. An interesting question is to understand whether
neural networks are able to exploit these symmetries and represent
these states efficiently. We describe long-range Hamiltonian dynamics
in the Methods section.

\begin{figure*}[th]
\centering
\subfloat[]{\includegraphics[width=0.4\textwidth, keepaspectratio]{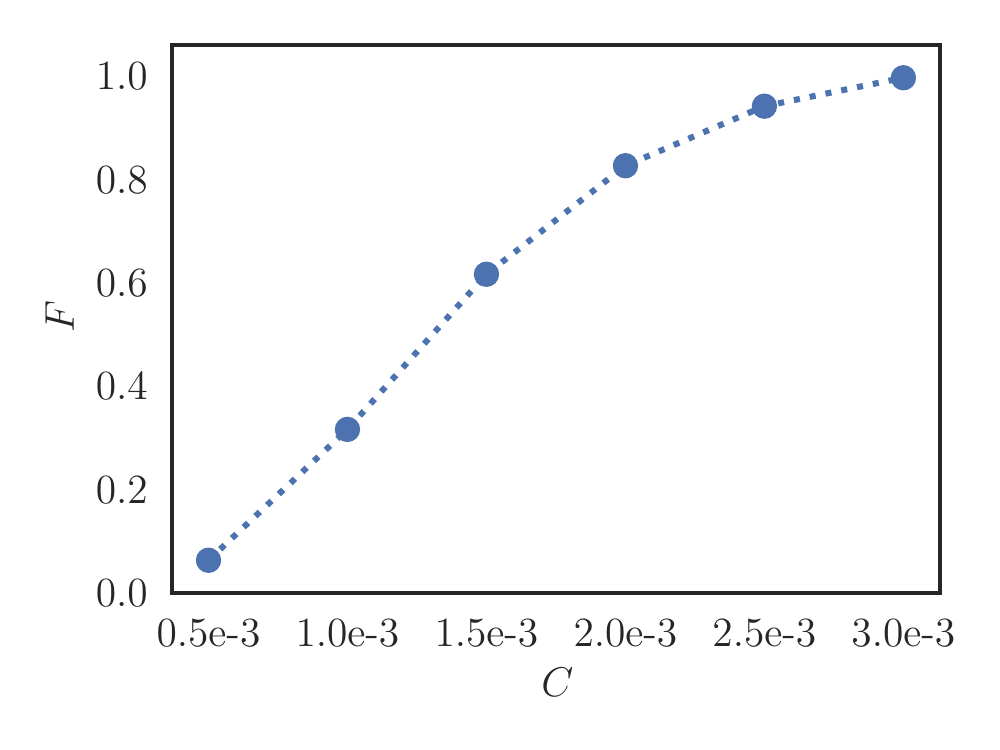}\label{fig:overlap_v_compression_evolve}}
\subfloat[]{\includegraphics[width=0.4\textwidth, keepaspectratio]{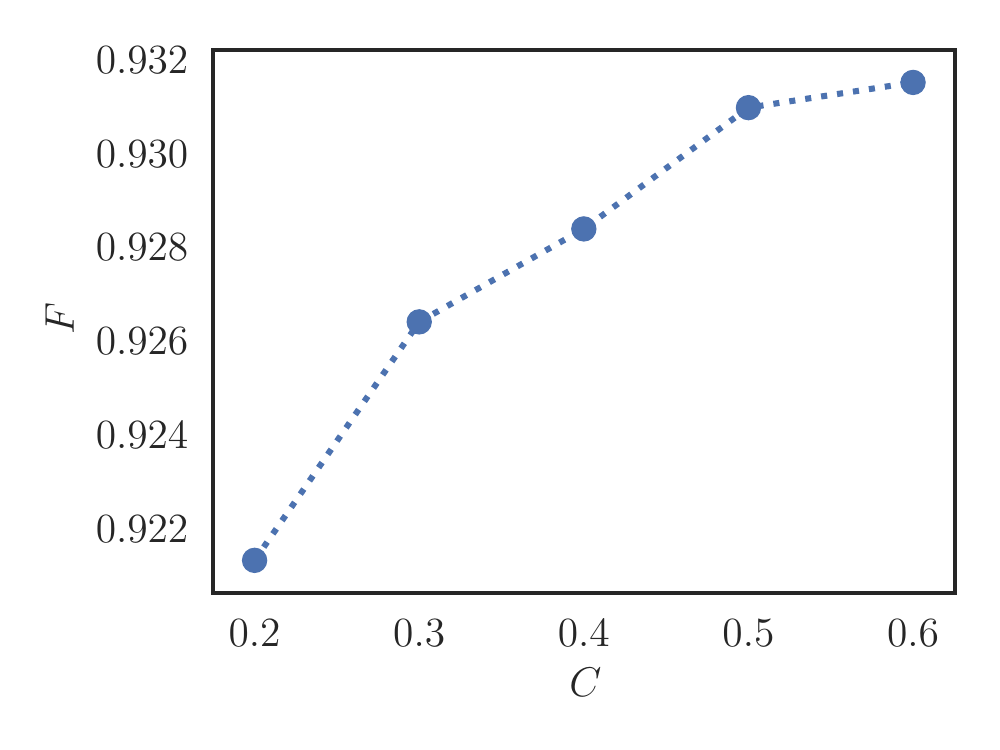}\label{fig:overlap_v_compression_hard}}

\caption{\textbf{VAEs can learn efficient representation of easy states and
can be used to characterize hard states.} Fidelity as a function of
compression $C=m/2^{n}$ for (a) an $18$-qubit state generated by
evolving $2^{-n/2}\sum_{i}\ket{i}$ using the long-range Hamiltonian
time evolution described in the Methods section for a time $t=20$
and (b) an $18$-qubit hard state generated according to~\cite{fefferman2015power}.
Figure (a) shows that the VAE can learn to represent efficiently with
almost perfect accuracy easy states that are challenging for MPS.
Figure (b) shows that hard quantum states can be compressed with high
reconstruction accuracy up to a factor $5$. The decoder in (a) has
$1$ hidden layer to allow for greater compression without incurring
in the saturation effects discussed in Methods section. The decoder
in (b) has $6$ hidden layers in order to maximise the representational
capability of the network.\label{fig:overlap_v_compression} }
\end{figure*}

Results are displayed in Fig.~\ref{fig:overlap_v_compression}. For
states obtained through Hamiltonian evolution we achieve with almost
maximum reconstruction accuracy compression levels of up to $C \approx  10^{-3}$.
This corresponds to a number of parameters $m=\mathcal{O}(100)\ll2^{18}$
which implies that the VAE has learned an efficient representation
of the state.

In the case of hard state we can reach a compression of $0.2$,
corresponding to a factor $5$ reduction in the number of parameters required to represent the state. Note that the entanglement properties of hard states \CC{are likely to} make them hard to compress for tensor network states. For example, \CC{if one wanted  to compress an $18$ qubits state using MPS (a type of tensor network that is known to be efficiently contractable)} we have found that the estimated bond dimension to reconstruct this state is $D \approx 460$. This number is obtained computing the largest bipartite entanglement entropy ($S$), and estimating the bond dimension with $D \approx 2^S$. Considering that an MPS has $D^2$ variational parameters (in the best case), this would yield about $200$ thousands variational parameters required to represent those hard states. The resulting MPS compressing factor is then about $1.23$, a significantly lower figure with respect to the $5$ compression factor obtained with VAEs. \CC{We note that this calculation only shows that  the entanglement structure of hard states is not well modelled by MPS. Other types of tensor networks might be more amenable to the specific structure of these states but it is unlikely these models will be computationally tractable.}

Although limited, the levels of compression we achieve for hard states could play a role in experiments aimed at showing quantum
supremacy. In this setting a quantum machine with a handful of noisy
qubits performs a task that is not reproducible even by the fastest
supercomputer. As recently highlighted by Montanaro and Harrow~\cite{harrow2017quantum}
one of the key challenges with quantum supremacy experiments is to
verify that the quantum machine is behaving as expected. Because quantum
computers are conjectured to not be efficiently simulatable, verifying
that a quantum machine is performing as expected is a hard problem
for classical machines. The paper by Jozsa and Strelchuk~\cite{jozsa2017efficient} provides an introduction
to several approaches to verification of quantum computation. Our
methods might allow to characterise the result of a computation by
reducing the complexity of the problem. \CC{Because any verification of quantum supremacy will likely involve a machine with only a few qubits above what can be efficiently classically simulated, even small reductions in the number of parameters of the state might allow to approximate relevant quantities in a computationally tractable way. Potentially, a neural network approach to verification}  can be accomplished by compressing a trusted initial state into a VAE whose parameters are then  evolved according to a set of rules specified by the quantum circuit. By comparing the experimental distribution with the one sampled with the VAE it is then possible to determine whether the device is faulty. \CC{We remark that this type of verification protocol would only ``approximately verify'' the system because of the errors introduced during the compression phase.}

\section*{Discussion}

\label{sec:conclusions}

In this work we introduced VAEs, a type of deep, generative, neural
network, as way to encode the probability distribution of quantum
states. Our methods are completely unsupervised, \textit{i.e.} do
not require a labelled training set. By means of numerical simulations
we showed that deep networks can represent hard quantum states that
can be efficiently obtained by a quantum computer better than shallow
ones. On the other hand, for states that are hard and conjectured
to be not efficiently producible by quantum computers, depth does
not appear to play a role in increasing the reconstruction accuracy.
Our results suggest that neural networks are able to capture correlations
in states that are provably hard to sample from for classical computers
but not for quantum ones. As already pointed out in other works, this
might signal that states that can be produced efficiently by a quantum
computer have a structure that is well represented by a layered neural
network.

Through numerical experiments we showed that our methods have two
important features. First, they are capable of representing, using
fewer parameters, states that that are known to have efficient representation
but where other classical approaches struggle. Second, VAEs can compress
hard quantum states up to a constant factor. However low, this compression
level might enable to approximately verify quantum states of a size
expected on near future quantum computers.

Presently, our methods allow to encode only the probability distribution
of a quantum state. Future research should focus on developing VAE
architectures that allow to reconstruct the full set of amplitudes.
Other interesting directions involve finding methods to compute the
quantum evolution of the parameters of the network and investigating
whether the depth of a quantum circuit is related to the optimal depth
of a VAE learning its output states. Finally, it is interesting to
investigate how information is encoded in the latent layers of the
network. Such analysis might provide novel tools to understand the
information theoretic properties of a quantum system.

\section*{Methods}

\subsection*{Numerical experiments}

\label{app:numsim}

All our networks were trained using the tensorflow r1.3 framework
on a single NVIDIA K80 GPU. Training was performed using backpropagation
and the Adam optimiser with initial learning rate of $10^{-3}$~\cite{kingma2014adam}.
Leaky rectified linear units (LReLU) function were used on all hidden
layers with the leak set to $0.2$~\cite{maas2013rectifier}. Sigmoid
activation functions were used on the final layer.

%At the start of training the regularisation objective can override the reconstruction objective and lead to poor performance. We used a \textit{warm up} schedule on the regularisation objective by increasing a weight on the regularisation error from $0$ to $0.85$ linearly during training~\cite{sonderby2016ladder}.

\AR{Training involves optimising two objectives: the reconstruction loss and the regularization loss. We used a warm up schedule on the regularisation objective by increasing a weight on the regularisation loss from $0$ to $0.85$ linearly during training ~\cite{sonderby2016ladder}.  This turned out to be critical, especially for hard states. A consequence of this approach is that the model does not learn the distribution until close to the end of training irrespective of the number of training iterations. Each network was trained using $50,000$ batches of $1000$ samples each. Each sample consists of a binary string representing a measurement outcome.}

%Each network was trained for $50,000$ iterations with batches of $1000$ examples in each iteration. Examples are binary strings corresponding to measurement outcomes sampled randomly from the state vector being learned.

Following training the state was reconstructed from the VAE decoder
by drawing $100(2^{n})$ samples from a multivariate Gaussian with
zero mean and unit variance. The samples were decoded by the decoder
to generate measurement outcomes in the form of binary strings. The
relative frequency of each string was recorded and used to reconstruct
the learned distribution which was compared to the true distribution
to determine its fidelity.

In all experiments the number of nodes in the latent layer is the
same as the number of qubits. Using fewer or more nodes in this layer
resulted in worse performance. The number of nodes in the hidden layers
is determined by the number of layers and the compression $C$ defined
by $\frac{m}{2^{n}}$ where $n$ is the number of qubits and $m$
is the number of parameters in the decoder. In all cases the encoder
has the same number of hidden layers and nodes in each layer as the
decoder.

We compress the VAE representation of a quantum state by removing
neurons from each hidden layer of the VAE. For small $n$'s achieving
a high level of compression caused instabilities in the network (\textit{i.e.}
the reconstruction accuracy became more dependent on the weight initialisation).
In this respect we note that, by restricting the number of neurons
in the penultimate layer, we are effectively constraining the number
of possible basis states that can be expressed in the output layer
and, as a result, the number of configurations the VAE can sample
from. This can be shown noting that the activation functions of the
penultimate layer generate a set of linear inequalities that must
be simultaneously satisfied. A geometric argument that involves how
many regions of an $n$-dimensional space $m$ hyperplanes can separate
lead to conclude that, to have full expressive capability, the penultimate
layer must include at least $n$ neurons. Similar arguments have been
discussed in~\cite{huang1991bounds} for multilayer perceptrons.

%We estimate the quality of the VAE representation with the fidelity.
%Given a quantum state $\ket{\psi}$ and its VAE approximation $\ket{\mathrm{VAE}}$
%we can write the distance between $\ket{\psi}$ and $\ket{\mathrm{VAE}}$
%as $F(\psi,\mathrm{VAE})=|\langle\psi|\mathrm{VAE}\rangle|$. When
%considering the probability distributions $p_{\psi}(i)$ and $p_{\mathrm{VAE}}(i)$
%corresponding to the quantum states, we get that the fidelity corresponds
%to the Bhattacharyya coefficient, \textit{i.e.} $F(\psi,\mathrm{VAE})=\sum_{i}\sqrt{p_{\psi}(i)p_{\mathrm{VAE}}(i)}$.
%This distance tells us how much two states overlap. When $F=1$ the
%two states are indistinguishable.

\subsection*{States that are classically hard to sample from}

\label{app:hardgen}

We study the learnability of a special class of hard states introduced
by Fefferman and Umans~\cite{fefferman2015power} which is produced
by a certain quantum computational processes which exhibit quantum
``supremacy''. The latter is a phenomenon whereby a quantum circuit
which consists of quantum gates and measurements on a constant number
of qubit lines samples from a particular class of distributions which
is known to be hard to sample from on a classical computer modulo
some very plausible computational complexity assumptions. To demonstrate
quantum supremacy one only requires quantum gates to operate within
a certain fidelity without full error-correction. This makes efficient
sampling from such distributions feasible to execute on near-term
quantum devices and opens the search for possibilities to look for
practically-relevant decision problems.

To construct a distribution one starts from an encoding function $h:[m]\to\{0,1\}^{N}$.
The function $h$ performs an efficient encoding of its argument and
is used to construct the following so-called efficiently specifiable
polynomial on $n$ variables: 
\begin{equation}
Q(X_{1},\dots,X_{N})=\sum_{z\in[m]}X_{1}^{h(z)_{1}}\dots X_{N}^{h(z)_{N}},
\end{equation}
where $h(z)_{i}$ means that we take only the $i$-th bit, and $m$
is an arbitrary integer. In the following, we pick $h$ to be related
to the permanent. More specifically, $h:[0,n!-1]\to\{0,1\}^{n^{2}}$
maps the $i$-th permutation (out of $n!$) to a string which encodes
its $n\times n$ permutation matrix in a natural way resulting in
a $N$-coordinate vector, where $N=n^{2}$. To encode a number $A\in[0,n!-1]$
in terms of its permutation vector we first represent $A$ in factorial
number system to get $A'$ obtaining the $N$-coordinate vector which
identifies a particular permutation $\sigma$.

With the above encoding, our efficiently specifiable polynomial $Q$
will have the form:

\begin{equation}
Q(X_{1},\dots,X_{N})=\sum_{z\in[n!-1]}X_{1}^{h(z)_{1}}\dots X_{N}^{h(z)_{N}}.
\end{equation}

Fix some number $L$ and consider the following set of vectors $y=(y_{1},\ldots,y_{N})\in[0,L-1]^{N}$
(i.e. each $y_{j}$ ranges between $0$ and $L-1$). For each $y$
construct another vector $Z_{y}=(z_{y_{1}},\dots,z_{y_{N}})$ constructed
as follows: each $z_{y_{j}}$ corresponds to a complex $L$-ary root
of unity raised to power $y_{j}$. For instance, pick $L=4$ and consider
$y'=(1,2,3,0,2,3,0,4)$. Then the corresponding vector $Z_{y'}=(w^{1},w^{2},w^{3},w^{0},w^{2},w^{3},w^{0},w^{4})$,
where $w=e^{2\pi i/4}$ (for an arbitrary $L$ it will be $e^{2\pi i/L}$).

Having defined $Q$ fixed $L$ we are now ready to construct each
element of the ``hard'' distribution ${\cal D}_{Q,L}$:

\begin{equation}
\text{Pr}_{{\cal D}_{Q,L}}[y]=\frac{|Q(Z_{y})|^{2}}{L^{N}n!}.
\end{equation}

A quantum circuit which performs sampling is remarkably easy. It amounts
to applying the quantum Fourier transform to a uniform superposition
which was transformed by $h$ and measuring in the standard basis
(see Theorem 4 of Section 4 of ~\cite{fefferman2015power}).

Classical sampling of distributions based on the above efficiently
specifiable polynomial is believed to be hard in particular because
it contains the permanent problem. Thus, the existence of an efficient
classical sampler would imply a collapse of the Polynomial Hierarchy
to the third level (see Section 5 and 6 of~\cite{fefferman2015power}
for detailed proof).

\subsection*{Long-range quantum Hamiltonians}

\label{app:hamil}

The long-range Hamiltonian we consider has the form: 
\begin{equation}
\left|\Psi(t)\right\rangle =e^{-i\mathcal{H}t}|\Psi(t=0)\rangle,
\end{equation}
where 
\begin{equation}
\mathcal{H}=\sum_{i<j}V(i,j)\left(\sigma_{i}^{x}\sigma_{j}^{x}+\sigma_{i}^{y}\sigma_{j}^{y}\right),
\end{equation}
and $V(i,j)=1/|i-j|^{3/4}$ is a long-range two-body interaction,
and the initial state is a fully polarized state is the product state
$|\Psi(t=0)\rangle=2^{-n/2}\sum_{i}\ket{i}$. At long propagation
times $t\gg1$, the resulting states are highly entangled, and are
for example, challenging for MPS-based tomography~\cite{cramer2010efficient}.
To assess the ability of VAE to compress highly entangled states,
we focus on the task of reconstructing the outcomes of experimental
measurements in the computational basis. In particular, we generate
samples distributed according to the probability density $|\Psi_{i}(t)|^{2}$,
and reconstruct this distribution with our generative, deep models.

\paragraph*{Acknowledgements. }

We thank Carlo Ciliberto, Danial Dervovic, Alessandro Davide Ialongo,
Joshua Lockhart, and Gillian Marshall for helpful comments and discussions.
Andrea Rocchetto is supported by an EPSRC DTP Scholarship and by QinetiQ.
Edward Grant is supported by EPSRC {[}EP/P510270/1{]}. Giuseppe Carleo
is supported by the European Research Council through the ERC Advanced
Grant SIMCOFE, and by the Swiss National Science Foundation through
NCCR QSIT. Sergii Strelchuk is supported by a Leverhulme Trust Early
Career Fellowship. Simone Severini is supported by The Royal Society,
EPSRC and the National Natural Science Foundation of China.

\paragraph*{Contributions.}

The concept of using VAEs to encode probability distributions of quantum
states was conceived by A.R., E.G., and G.C. The complexity framework
was developed by A.R., G.C., and S.St. E.G. wrote the code and performed
the simulations with help from S.St. The project was supervised by
A.R. and S.Se. The first draft of the manuscript was prepared by A.R.
and all authors contributed to the writing of the final version. A.R. and E.G. contributed equally to this work.

\paragraph*{Competing Interests.}

The authors declare no competing financial interests.

\paragraph*{Data availability statements.}

All data needed to evaluate the conclusions are available from the
corresponding author upon reasonable request.

\bibliographystyle{naturemag}
\bibliography{bibliography_qVAE}

\begin{thebibliography}{10}
\expandafter\ifx\csname url\endcsname\relax
  \def\url#1{\texttt{#1}}\fi
\expandafter\ifx\csname urlprefix\endcsname\relax\def\urlprefix{URL }\fi
\providecommand{\bibinfo}[2]{#2}
\providecommand{\eprint}[2][]{\url{#2}}

\bibitem{nightingale1998quantum}
\bibinfo{author}{Nightingale, M.~P.} \& \bibinfo{author}{Umrigar, C.~J.}
\newblock \emph{\bibinfo{title}{Quantum Monte Carlo methods in physics and
  chemistry}}.
\newblock \bibinfo{number}{525} (\bibinfo{publisher}{Springer Science \&
  Business Media}, \bibinfo{year}{1998}).

\bibitem{gubernatis2016quantum}
\bibinfo{author}{Gubernatis, J.}, \bibinfo{author}{Kawashima, N.} \&
  \bibinfo{author}{Werner, P.}
\newblock \emph{\bibinfo{title}{Quantum Monte Carlo Methods}}
  (\bibinfo{publisher}{Cambridge University Press}, \bibinfo{year}{2016}).

\bibitem{suzuki1993quantum}
\bibinfo{author}{Suzuki, M.}
\newblock \emph{\bibinfo{title}{Quantum Monte Carlo methods in condensed matter
  physics}} (\bibinfo{publisher}{World scientific}, \bibinfo{year}{1993}).

\bibitem{verstraete2008matrix}
\bibinfo{author}{Verstraete, F.}, \bibinfo{author}{Murg, V.} \&
  \bibinfo{author}{Cirac, J.~I.}
\newblock \bibinfo{title}{Matrix product states, projected entangled pair
  states, and variational renormalization group methods for quantum spin
  systems}.
\newblock \emph{\bibinfo{journal}{Advances in Physics}}
  \textbf{\bibinfo{volume}{57}}, \bibinfo{pages}{143--224}
  (\bibinfo{year}{2008}).

\bibitem{orus2014practical}
\bibinfo{author}{Or{\'u}s, R.}
\newblock \bibinfo{title}{A practical introduction to tensor networks: Matrix
  product states and projected entangled pair states}.
\newblock \emph{\bibinfo{journal}{Annals of Physics}}
  \textbf{\bibinfo{volume}{349}}, \bibinfo{pages}{117--158}
  (\bibinfo{year}{2014}).

\bibitem{carleo2017solving}
\bibinfo{author}{Carleo, G.} \& \bibinfo{author}{Troyer, M.}
\newblock \bibinfo{title}{Solving the quantum many-body problem with artificial
  neural networks}.
\newblock \emph{\bibinfo{journal}{Science}} \textbf{\bibinfo{volume}{355}},
  \bibinfo{pages}{602--606} (\bibinfo{year}{2017}).

\bibitem{deng_quantum_2017}
\bibinfo{author}{Deng, D.-L.}, \bibinfo{author}{Li, X.} \&
  \bibinfo{author}{Das~Sarma, S.}
\newblock \bibinfo{title}{Quantum {Entanglement} in {Neural} {Network}
  {States}}.
\newblock \emph{\bibinfo{journal}{Physical Review X}}
  \textbf{\bibinfo{volume}{7}}, \bibinfo{pages}{021021} (\bibinfo{year}{2017}).

\bibitem{nomura_restricted-boltzmann-machine_2017}
\bibinfo{author}{Nomura, Y.}, \bibinfo{author}{Darmawan, A.},
  \bibinfo{author}{Yamaji, Y.} \& \bibinfo{author}{Imada, M.}
\newblock \bibinfo{title}{Restricted-{Boltzmann}-{Machine} {Learning} for
  {Solving} {Strongly} {Correlated} {Quantum} {Systems}}.
\newblock \emph{\bibinfo{journal}{arXiv:1709.06475}}  (\bibinfo{year}{2017}).

\bibitem{deng_exact_2016}
\bibinfo{author}{Deng, D.-L.}, \bibinfo{author}{Li, X.} \&
  \bibinfo{author}{Sarma, S.~D.}
\newblock \bibinfo{title}{Exact {Machine} {Learning} {Topological} {States}}.
\newblock \emph{\bibinfo{journal}{arXiv:1609.09060}}  (\bibinfo{year}{2016}).

\bibitem{glasser_neural_2017}
\bibinfo{author}{Glasser, I.}, \bibinfo{author}{Pancotti, N.},
  \bibinfo{author}{August, M.}, \bibinfo{author}{Rodriguez, I.~D.} \&
  \bibinfo{author}{Cirac, J.~I.}
\newblock \bibinfo{title}{Neural {Networks} {Quantum} {States}, {String}-{Bond}
  {States} and chiral topological states}.
\newblock \emph{\bibinfo{journal}{arXiv:1710.04045}}  (\bibinfo{year}{2017}).

\bibitem{kaubruegger_chiral_2017}
\bibinfo{author}{Kaubruegger, R.}, \bibinfo{author}{Pastori, L.} \&
  \bibinfo{author}{Budich, J.~C.}
\newblock \bibinfo{title}{Chiral {Topological} {Phases} from {Artificial}
  {Neural} {Networks}}.
\newblock \emph{\bibinfo{journal}{arXiv:1710.04713}}  (\bibinfo{year}{2017}).

\bibitem{torlai2017many}
\bibinfo{author}{Torlai, G.} \emph{et~al.}
\newblock \bibinfo{title}{Many-body quantum state tomography with neural
  networks}.
\newblock \emph{\bibinfo{journal}{arXiv preprint arXiv:1703.05334}}
  (\bibinfo{year}{2017}).

\bibitem{perez2006matrix}
\bibinfo{author}{Perez-Garcia, D.}, \bibinfo{author}{Verstraete, F.},
  \bibinfo{author}{Wolf, M.~M.} \& \bibinfo{author}{Cirac, J.~I.}
\newblock \bibinfo{title}{Matrix product state representations}.
\newblock \emph{\bibinfo{journal}{arXiv preprint quant-ph/0608197}}
  (\bibinfo{year}{2006}).

\bibitem{gao2017efficient}
\bibinfo{author}{Gao, X.} \& \bibinfo{author}{Duan, L.-M.}
\newblock \bibinfo{title}{Efficient representation of quantum many-body states
  with deep neural networks}.
\newblock \emph{\bibinfo{journal}{Nature Communications}}
  \textbf{\bibinfo{volume}{8}}, \bibinfo{pages}{662} (\bibinfo{year}{2017}).

\bibitem{chen_equivalence_2017}
\bibinfo{author}{Chen, J.}, \bibinfo{author}{Cheng, S.}, \bibinfo{author}{Xie,
  H.}, \bibinfo{author}{Wang, L.} \& \bibinfo{author}{Xiang, T.}
\newblock \bibinfo{title}{On the {Equivalence} of {Restricted} {Boltzmann}
  {Machines} and {Tensor} {Network} {States}}.
\newblock \emph{\bibinfo{journal}{arXiv:1701.04831}}  (\bibinfo{year}{2017}).

\bibitem{huang_neural_2017}
\bibinfo{author}{Huang, Y.} \& \bibinfo{author}{Moore, J.~E.}
\newblock \bibinfo{title}{Neural network representation of tensor network and
  chiral states}.
\newblock \emph{\bibinfo{journal}{arXiv:1701.06246}}  (\bibinfo{year}{2017}).

\bibitem{clark_unifying_2017}
\bibinfo{author}{Clark, S.~R.}
\newblock \bibinfo{title}{Unifying {Neural}-network {Quantum} {States} and
  {Correlator} {Product} {States} via {Tensor} {Networks}}.
\newblock \emph{\bibinfo{journal}{arXiv:1710.03545}}  (\bibinfo{year}{2017}).

\bibitem{mhaskar2016learning}
\bibinfo{author}{Mhaskar, H.}, \bibinfo{author}{Liao, Q.} \&
  \bibinfo{author}{Poggio, T.}
\newblock \bibinfo{title}{Learning functions: When is deep better than
  shallow}.
\newblock \emph{\bibinfo{journal}{arXiv preprint arXiv:1603.00988}}
  (\bibinfo{year}{2016}).

\bibitem{telgarsky2016benefits}
\bibinfo{author}{Telgarsky, M.}
\newblock \bibinfo{title}{Benefits of depth in neural networks}.
\newblock \emph{\bibinfo{journal}{arXiv preprint arXiv:1602.04485}}
  (\bibinfo{year}{2016}).

\bibitem{eldan2016power}
\bibinfo{author}{Eldan, R.} \& \bibinfo{author}{Shamir, O.}
\newblock \bibinfo{title}{The power of depth for feedforward neural networks}.
\newblock In \emph{\bibinfo{booktitle}{Conference on Learning Theory}},
  \bibinfo{pages}{907--940} (\bibinfo{year}{2016}).

\bibitem{kingma2013auto}
\bibinfo{author}{Kingma, D.~P.} \& \bibinfo{author}{Welling, M.}
\newblock \bibinfo{title}{Auto-encoding variational bayes}.
\newblock \emph{\bibinfo{journal}{arXiv preprint arXiv:1312.6114}}
  (\bibinfo{year}{2013}).

\bibitem{boixo2016characterizing}
\bibinfo{author}{Boixo, S.} \emph{et~al.}
\newblock \bibinfo{title}{Characterizing quantum supremacy in near-term
  devices}.
\newblock \emph{\bibinfo{journal}{arXiv preprint arXiv:1608.00263}}
  (\bibinfo{year}{2016}).

\bibitem{fuchs1994ensemble}
\bibinfo{author}{Fuchs, C.~A.} \& \bibinfo{author}{Caves, C.~M.}
\newblock \bibinfo{title}{Ensemble-dependent bounds for accessible information
  in quantum mechanics}.
\newblock \emph{\bibinfo{journal}{Physical Review Letters}}
  \textbf{\bibinfo{volume}{73}}, \bibinfo{pages}{3047} (\bibinfo{year}{1994}).

\bibitem{fefferman2014power}
\bibinfo{author}{Fefferman, W.~J.}
\newblock \emph{\bibinfo{title}{The power of quantum Fourier sampling}}.
\newblock Ph.D. thesis, \bibinfo{school}{California Institute of Technology}
  (\bibinfo{year}{2014}).

\bibitem{aaronson2011computational}
\bibinfo{author}{Aaronson, S.} \& \bibinfo{author}{Arkhipov, A.}
\newblock \bibinfo{title}{The computational complexity of linear optics}.
\newblock In \emph{\bibinfo{booktitle}{Proceedings of the forty-third annual
  ACM symposium on Theory of computing}}, \bibinfo{pages}{333--342}
  (\bibinfo{organization}{ACM}, \bibinfo{year}{2011}).

\bibitem{morningstar2017deep}
\bibinfo{author}{Morningstar, A.} \& \bibinfo{author}{Melko, R.~G.}
\newblock \bibinfo{title}{Deep learning the ising model near criticality}.
\newblock \emph{\bibinfo{journal}{arXiv preprint arXiv:1708.04622}}
  (\bibinfo{year}{2017}).

\bibitem{levine2017deep}
\bibinfo{author}{Levine, Y.}, \bibinfo{author}{Yakira, D.},
  \bibinfo{author}{Cohen, N.} \& \bibinfo{author}{Shashua, A.}
\newblock \bibinfo{title}{Deep learning and quantum entanglement: Fundamental
  connections with implications to network design.}
\newblock \emph{\bibinfo{journal}{arXiv preprint arXiv:1704.01552}}
  (\bibinfo{year}{2017}).

\bibitem{fefferman2015power}
\bibinfo{author}{Fefferman, B.} \& \bibinfo{author}{Umans, C.}
\newblock \bibinfo{title}{The power of quantum fourier sampling}.
\newblock \emph{\bibinfo{journal}{arXiv preprint arXiv:1507.05592}}
  (\bibinfo{year}{2015}).

\bibitem{richerme2014non}
\bibinfo{author}{Richerme, P.} \emph{et~al.}
\newblock \bibinfo{title}{Non-local propagation of correlations in quantum
  systems with long-range interactions}.
\newblock \emph{\bibinfo{journal}{Nature}} \textbf{\bibinfo{volume}{511}},
  \bibinfo{pages}{198--201} (\bibinfo{year}{2014}).

\bibitem{harrow2017quantum}
\bibinfo{author}{Harrow, A.~W.} \& \bibinfo{author}{Montanaro, A.}
\newblock \bibinfo{title}{Quantum computational supremacy}.
\newblock \emph{\bibinfo{journal}{Nature}} \textbf{\bibinfo{volume}{549}},
  \bibinfo{pages}{203--209} (\bibinfo{year}{2017}).

\bibitem{jozsa2017efficient}
\bibinfo{author}{Jozsa, R.} \& \bibinfo{author}{Strelchuk, S.}
\newblock \bibinfo{title}{{Efficient classical verification of quantum
  computations}}.
\newblock \emph{\bibinfo{journal}{arXiv preprint arXiv:1705.02817}}
  (\bibinfo{year}{2017}).

\bibitem{kingma2014adam}
\bibinfo{author}{Kingma, D.} \& \bibinfo{author}{Ba, J.}
\newblock \bibinfo{title}{Adam: A method for stochastic optimization}.
\newblock \emph{\bibinfo{journal}{arXiv preprint arXiv:1412.6980}}
  (\bibinfo{year}{2014}).

\bibitem{maas2013rectifier}
\bibinfo{author}{Maas, A.~L.}, \bibinfo{author}{Hannun, A.~Y.} \&
  \bibinfo{author}{Ng, A.~Y.}
\newblock \bibinfo{title}{Rectifier nonlinearities improve neural network
  acoustic models}.
\newblock In \emph{\bibinfo{booktitle}{Proc. ICML}}, vol.~\bibinfo{volume}{30}
  (\bibinfo{year}{2013}).

\bibitem{sonderby2016ladder}
\bibinfo{author}{S{\o}nderby, C.~K.}, \bibinfo{author}{Raiko, T.},
  \bibinfo{author}{Maal{\o}e, L.}, \bibinfo{author}{S{\o}nderby, S.~K.} \&
  \bibinfo{author}{Winther, O.}
\newblock \bibinfo{title}{Ladder variational autoencoders}.
\newblock In \emph{\bibinfo{booktitle}{Advances in Neural Information
  Processing Systems}}, \bibinfo{pages}{3738--3746} (\bibinfo{year}{2016}).

\bibitem{huang1991bounds}
\bibinfo{author}{Huang, S.-C.} \& \bibinfo{author}{Huang, Y.-F.}
\newblock \bibinfo{title}{Bounds on the number of hidden neurons in multilayer
  perceptrons}.
\newblock \emph{\bibinfo{journal}{IEEE transactions on neural networks}}
  \textbf{\bibinfo{volume}{2}}, \bibinfo{pages}{47--55} (\bibinfo{year}{1991}).

\bibitem{cramer2010efficient}
\bibinfo{author}{Cramer, M.} \emph{et~al.}
\newblock \bibinfo{title}{Efficient quantum state tomography}.
\newblock \emph{\bibinfo{journal}{Nature communications}}
  \textbf{\bibinfo{volume}{1}}, \bibinfo{pages}{149} (\bibinfo{year}{2010}).

\end{thebibliography}

\end{document}